\documentstyle[12pt,dina4p,epsfig]{article}
\title{
  \vskip-2cm
  {\baselineskip16pt
    \centerline{\normalsize \tt DESY 96-246 \hfill ISSN 0418-9833}
    \centerline{\normalsize \tt hep-ph/9611450 \hfill}
    \centerline{\normalsize \tt November 1996 \hfill}
  }
  \vskip2cm
  {\bf 
    Inclusive Two-Jet Production at HERA: \\
    Direct and Resolved Cross Sections in Next-to-Leading Order QCD
  }
  \author{
    {M.\ Klasen\footnote{Now at Deutsches Elektronen-Synchrotron (DESY),
     Notkestra\ss{}e 85, D-22607 Hamburg, Germany.}, G.\ Kramer} \\
    {II. Institut f\"ur Theoretische Physik}\thanks
    {Supported by Bundesministerium f\"ur Forschung und
     Technologie, Bonn, Germany under Contract 05\,7HH92P(0) and
     EEC Program "Human Capital and Mobility" through Network
     "Physics at High Energy Colliders" under Contract
     CHRX-CT93-0357 (DG12 COMA)} \\
     {Universit\"at Hamburg} \\
     {D - 22761 Hamburg, Germany}
    }
  \date{}
}
\begin{document}
\maketitle
\vspace{3cm}
\begin{abstract}
\thispagestyle{empty}
We have calculated inclusive two-jet cross sections in next-to-leading order
QCD for low $Q^2$ $ep$ collisions superimposing direct and resolved
contributions. Infrared and collinear singularities in the virtual and real
contributions are cancelled with the phase space slicing method. Various
inclusive two-jet distributions have been computed. The results are compared
with recent data from the ZEUS collaboration at HERA.
\end{abstract}
\newpage
\setcounter{page}{1}
\section{Introduction}
The analysis of large transverse energy ($E_T$) jets produced in $ep$
scattering with very small virtuality ($Q^2\simeq 0$) offers the possibility
to learn about the partonic structure of real photons. The hard momentum
scale $E_T$ allows the application of perturbative QCD to predict cross
sections for the production of two or more high-$E_T$ jets. \\

As is well known, in lowest order (LO) QCD two distinct processes contribute
to the hard scattering cross sections. In the direct process, the incoming
photon interacts in a point-like fashion with the quark from the proton via
the QCD Compton scattering $\gamma q\rightarrow gq$ or via the photon-gluon
fusion process $\gamma g\rightarrow q\overline{q}$, if a gluon comes out
of the proton. In the resolved process, the photon acts as a source of partons,
which can scatter off the partons in the proton. Therefore, the cross
section for the resolved process depends on the parton distributions in the
photon in addition to the parton distribution functions in the proton. \\

The two processes have distinctly different event structures. Whereas the
direct process results in events with a characteristic (2+1)-jet structure,
i.e.~two high-$E_T$ jets and one low-$E_T$ remnant jet from the proton, the
resolved photon events have a characteristic (2+2)-jet structure, where in
addition to the two high-$E_T$ jets and the low-$E_T$ proton remnant a second
jet of low-$E_T$ fragments of the photon is produced. While in the (2+1)-jet
events the total photon energy contributes to the hard scattering process,
only a fraction participates in the (2+2)-jet events. \\

This characterization of the direct and resolved processes is valid only in LO.
In next-to-leading order (NLO), the direct cross section may also have
contributions with a photon remnant jet. In addition, both components are
related to each other through the factorization scheme and scale at the photon
leg. The dependence of the NLO direct cross section on this scheme/scale
must cancel up to NNLO terms against the scheme/scale dependence of the
resolved cross section. Therefore in NLO both components must be considered
together and consistently be calculated up to NLO in the parton distributions
of the photon and the proton, respectively, and in the hard scattering cross
sections. \\

The simplest observable is the inclusive single-jet cross section. It depends
on the transverse energy and the rapidity of the observed jet. Complete
NLO calculations for this cross section have been done previously
\cite{xxx1,xxx2,xxx3} and have been
compared to experimental data from ZEUS \cite{xxx4}
and H1 \cite{xxx5}. In these calculations special techniques for cancelling
soft and collinear singularities have been applied, which are suitable only
for the computation of inclusive single-jet cross sections. To obtain different
cross sections, as for example the inclusive two-jet cross section, the whole
calculation must be repeated. Inclusive two-jet cross sections depend on one
more variable. Therefore they are a much more stringent test of the QCD
predictions than inclusive single-jet cross sections. Furthermore, for
a comparison with experimental data the calculations must be performed in
such a way that the experimental cuts can be built in easily. For this
purpose, the so-called phase space slicing method is suited particularly well.
In this method, an invariant mass resolution cut is introduced to isolate
the soft and collinear singularities of initial and final states. This
resolution cut is purely technical and is chosen small enough not to disturb
any experimental constraints or jet recombination requirements. Based on this
method we recently calculated the inclusive two-jet cross section for
direct photoproduction \cite{xxx6}. The results of this calculation were
compared with data from the ZEUS collaboration \cite{xxx7}, in which the direct
contribution and the resolved contribution were separated with a cut on
$x_\gamma$, i.e.~the momentum fraction of the photon entering the hard
scattering. The contribution of the resolved photon in the enriched direct
$\gamma$ sample was estimated in lowest order since it was supposed to
contribute only a fraction in this sample. As a result of this investigation
it turned out that the resolved contribution in the enriched direct $\gamma$
sample was not negligible and could be, depending on the assumed photon
structure function, as large as 30\% of the total two-jet cross section.
It is clear that the estimate of the resolved
cross section in this region by a LO computation is insufficient and that
the NLO corrections of the resolved cross section must be included.
In addition, in the same analysis of the ZEUS two-jet data \cite{xxx7} the
inclusive two-jet cross section in an enriched resolved $\gamma$ sample
was measured, which could be a decisive test of the available parton
distributions of the photon. For this reason and since we expect more
two-jet experimental data in the future, a complete NLO calculation of the
inclusive two-jet cross section is mandatory. For the direct component,
we rely on our previous work \cite{xxx6,xxx8}. The NLO calculation of the
resolved component has been completed recently \cite{xxx9}. A detailed
description of the calculation will be presented in a separate publication
\cite{xxx10}. Here we shall only give a short presentation, report some results
on various inclusive two-jet cross sections, which might be measured in
the future, and recalculate the two-jet cross section in the enriched
direct $\gamma$ sample and the enriched resolved $\gamma$ sample
for a comparison with recent ZEUS data \cite{xxx11}. \\

The outline of the paper is as follows: in section 2, we explain the formalism
used to calculate the inclusive two-jet cross section for the resolved part.
In section 3, we present some results in order to demonstrate the flexibility
of our method and a comparison with recent ZEUS measurements.
Section 4 contains a summary.

\section{Next-To-Leading Order Cross Sections}
\subsection{Photon Spectrum}
Before we explain how the NLO corrections have been calculated, we must specify
the relation between the $ep$ and the $\gamma p$ cross sections. Let us start
with the electroproduction process
\begin{equation}
 e(k) + p(p) \rightarrow e'(k') + X,
\end{equation}
where $k$, $k'$, and $p$ are the four-momenta of the incoming and outgoing
electron (positron) and the proton, respectively. $X$ denotes a generic
hadronic system which will be specified later. $q=k-k'$ is the momentum
transfer of the electron to the photon with virtuality $Q^2=-q^2\simeq 0$.
For small $Q^2$, the $ep$ cross section can be calculated in the
Weizs\"acker-Williams approximation, where the $\gamma p$ cross section can
be factorized and the photon spectrum is given by the function
\begin{equation}
 F_{\gamma/e} (x_a) = \frac{\alpha}{2\pi}\frac{1+(1-x_a)^2}{x_a}
 \ln\frac{Q_{\max}^2(1-x_a)}{m_e^2x_a^2}.
\end{equation}
Here, $m_e$ is the electron mass, and $x_a=(pq)/(pk)\simeq E_{\gamma}/E_e$
is the fraction of the initial electron energy transferred to the photon.
The cross section for the process (1) $ep\rightarrow e'X$ is then given by
the convolution
\begin{equation}
 \mbox{d}\sigma (ep\rightarrow e'X) = \int\limits_{x_{a,\min}}^{x_{a,\max}}
 \mbox{d}x_a F_{\gamma/e}(x_a)\mbox{d}\sigma (\gamma p\rightarrow X).
\end{equation}
$Q_{\max}^2$, $x_{a,\max}\leq 1$, and $x_{a,\min}$ are determined by the
tagging conditions of the experiments at HERA and will be specified when
we present our numerical results. In (3), d$\sigma (\gamma p\rightarrow X)$
is the photoproduction cross section with initial photon energy $E_{\gamma}
=x_aE_e$ and photon virtuality $Q^2\simeq 0$.

\subsection{Jet Cross Sections in NLO}
As we explained in the introduction, inclusive jet cross sections d$\sigma
(\gamma p\rightarrow \mbox{jets})$ receive contributions from two components,
direct and resolved, depending on the way in which the incoming photons
participate in the hard scattering subprocesses. In the resolved case,
the NLO inclusive two-jet cross section is calculated from the
formula
\begin{eqnarray}
 \frac{\mbox{d}^3\sigma}{\mbox{d}E_T\mbox{d}\eta_1\mbox{d}\eta_2} &=&
 \sum_{a,b=q,g}\int\mbox{d}y_a\int\mbox{d}x_b F_{a/\gamma}(y_a,M_{\gamma}^2)
 F_{b/p}(x_b,M_p^2)\nonumber\\
 && 
 \left\{\left( \frac{\mbox{d}^3\sigma (ab\rightarrow \mbox{jet}_1+ \mbox{jet}_2
 )}{\mbox{d}E_T\mbox{d}\eta_1\mbox{d}\eta_2}\right)_{\mbox{LO}}
 +\frac{\alpha_s(\mu)}{2\pi}K_{ab}(R,M_{\gamma},M_p,\mu)\right\} .
\end{eqnarray}
In (4), $E_T$ is the transverse energy of the measured or trigger jet with
rapidity $\eta_1$. $\eta_2$ denotes the rapidity of another jet such that
in the three-jet sample these two jets have the largest and second largest
$E_T$, i.e.~$E_{T_1},E_{T_2} > E_{T_3}$. The first term on the right-hand
side of (4) is the LO cross section, and the second term is the NLO correction.
The latter depends on the factorization scales $M_{\gamma}$ and $M_p$ at the
photon and proton vertex, at which the initial state singularities are 
absorbed into the parton distribution functions of the photon and the proton.
$\mu$ is the renormalization scale. The variable $R$ is the usual cone size
parameter, which defines the size of the jets with transverse energy $E_T$,
rapidity $\eta$, and azimuthal angle $\phi$. When two partons fulfill the
Snowmass condition \cite{xxx12} with the cone size parameter $R$, they are
combined into one jet. The same jet definition must be used in the
analysis of the experimental data, if a meaningful comparison between theory
and experiment is intended. It is clear that either of the two jets can
consist of two partons inside a cone with radius $R$. Essentially the same
formula applies for the direct cross section. In this case, the photon
structure function $F_{a/\gamma}(y_a,M_{\gamma}^2)$ is replaced by
$\delta (1-y_a)$. \\

The NLO corrections $K_{ab}$ in (4) for the hard scattering subprocesses
are calculated with the phase space slicing method using an invariant mass
cut for the separation of the cross section for $\gamma p\rightarrow 2$ jets
and $\gamma p\rightarrow 3$ jets. This invariant mass cut is defined by
$2p_ip_j < ys$, where $s$ is the partonic center-of-mass energy
squared. So, for example, in the resolved subprocess $q_i
\overline{q}_j\rightarrow q_i\overline{q}_jg$, where $q_i$, $q_j$ can be
identical or non-identical quarks, the cross section has soft, initial, and
final state collinear singularities. The cross section d$\sigma (q_i
\overline{q}_j\rightarrow q_i\overline{q}_jg)$ is integrated over these
singular regions up to the cut-off $y$, which isolates the respective
singularities in terms of poles in $\varepsilon=(4-d)/2$, where $d$ is the
dimension in the dimensional regularization method. These singularities
cancel against the singular contributions of the virtual corrections to
$q_i\overline{q}_j\rightarrow q_i\overline{q}_j$ and against the
subtraction terms at the scales $M_{\gamma}$ and $M_p$, which are absorbed
into the parton distribution functions of the photon and the proton,
respectively. Outside the cut-off region, i.e.~for $2p_ip_j > ys$, we have
genuine $q_i\overline{q}_jg$ final states. The cross section in this region
is subdivided into the two-jet cross section and the three-jet cross section
contribution, depending on whether two of the final state partons are combined
according to the Snowmass condition or not. The LO contribution, the
NLO virtual contributions, and the NLO corrections inside the $y$-cut
contribute to the two-jet cross section together with the contributions
inside the recombination cone with radius $R$. The part of the $q_i
\overline{q}_jg$ cross section not fulfilling the cone recombination condition
is the three-jet cross section, from which we have calculated the inclusive
two-jet cross section as a function of $E_T$, $\eta_1$, and $\eta_2$.
For the exclusive two-jet cross section, $E_{T_1}=E_{T_2}=E_T$. To separate
double singularities in the $q_i\overline{q}_jg$ cross section part we
have used the method of partial fractioning. The same partial fractioned
expressions are applied for the calculation inside and outside the $y$-cut.
Inside the $y$-cut region, the integrations are done analytically with
the approximation that terms ${\cal O} (y)$ are neglected. This is
necessary since otherwise with a $d$-dimensional space-time the necessary
integration could not be carried out. Outside the $y$-cut region, the
contributions to the inclusive cross section are evaluated with no further
approximations. Because of the approximation in the analytic part inside
the $y$-cut, the parameter $y$ must be chosen very small. It turned out
that for $y$ of the order $10^{-3}$ to $10^{-4}$ the inclusive cross section
is independent of the cut-off parameter $y$. However, the inclusive
two-jet cross section depends on the cone size $R$, which must be chosen
in accordance with the analysis of the experimental data. In the case of all
the other hard subprocesses for the resolved cross section, of which there
are many and which will not be listed here, the calculation proceeds in the
same way. The complete list will be given in a separate publication
\cite{xxx10} and can also be found in \cite{xxx9,xxx13}. In these
references, the details of the analytic integrations and the cancellation
of the singularities for NLO corrections is described as well. For the
direct cross section, the calculation of the NLO correction was already
reported and is given with all details in our previous work \cite{xxx6,xxx8,
xxx9}. \\

Before the final results were obtained, some tests of the NLO corrections
have been performed. For similar tests of the NLO corrections of the direct
cross section see \cite{xxx6}. The same tests were done for the resolved
subprocesses. So we checked that the cross sections are independent of the
cut-off $y$ if $y$ is chosen small enough. This was the case for $y\leq 
10^{-3}$
in all considered cases under different kinematical conditions. For
$y>10^{-3}$, we observed some small $y$ dependence caused by the
approximations in the analytical integrations. Second, the resolved inclusive
one-jet cross section was calculated and compared with earlier results for
which a completely different method for cancelling the infrared and collinear
singularities was applied \cite{xxx2,xxx14}. Very good agreement was found
in all channels and for the total resolved cross section. Third we tested
that the sum of the NLO direct and the LO resolved cross section is independent
of the factorization scale $M_{\gamma}$ similar to such a test for the
single-jet inclusive cross section earlier \cite{xxx15}. Details of these
tests can be found in \cite{xxx9}. \\

For the calculation of the inclusive jet cross sections, we need the parton
distributions $F_{a/\gamma}$ of the photon and $F_{b/p}$ of the proton,
respectively (see (4)). We have chosen the NLO set of Gl\"uck, Reya, and
Vogt (GRV) in the $\overline{\mbox{MS}}$ scheme for $F_{a/\gamma}$
\cite{xxx16} and the CTEQ3M set, i.e.~also a $\overline{\mbox{MS}}$ set in
NLO, for $F_{b/p}$ \cite{xxx17}. All scales are set to $\mu=M_{\gamma}=
M_p=E_T$. $\alpha_s$ is calculated from the two-loop formula with $n_f=4$
massless flavors with $\Lambda_{\overline{\rm MS}}^{(4)} = 239$
MeV, equal to the $\Lambda$-value of the CTEQ3M parton distributions. The
charm quarks are treated as light flavors with the boundary condition
that the charm content of the two structure functions vanishes for
$M_{\gamma}^2$, $M_p^2 < m_c^2$ with $m_c =1.5$ GeV. 

\section{Results and Comparison with ZEUS Data}
In this section we present some characteristic numerical results for
two-jet cross sections concentrating on the resolved contribution. All
results are obtained with the phase space slicing method. Except where we
compare with recent ZEUS data, we do not apply special cuts on
kinematical variables of the initial or final states dictated by the
experimental analysis, although this is necessary if one wants to confront
the prediction with experimental data. Therefore, the photon energy fraction
$x_a$ in (2) is allowed to vary in the kinematically allowed range from
$x_{a,\min}\leq x_a\leq 1$. $Q_{\max}^2 = 4$ GeV$^2$ in (2) is the same as
usually used in the jet analysis of the ZEUS experiment. \\

The further evaluation of the two-jet cross section is based on two separate
contributions -- a set of two-body contributions and a set of three body
contributions. Each set is completely finite, as all singularities have been
cancelled or absorbed into structure functions in the same way as in the
case of the direct cross section \cite{xxx8}. Each part depends separately on
the cut-off $y$. The analytic contributions are valid only for very small $y$,
since terms of ${\cal O} (y)$ have been neglected. For very small $y$, the
two contributions have no physical meaning. In this case, the terms
depending on $\ln y$ force the two-body contributions to become negative,
whereas the three-body cross sections are large and positive. The resolved
cross section shows a similar behavior \cite{xxx9}. When the two-
and three-body contributions are superimposed to yield the inclusive two-jet
cross section, the dependence on the cut-off will cancel. This will only
be the case if the inclusive cross sections are defined in the correct way,
so that they are infrared safe. We have checked this explicitly by
varying $y$ in the superposition of the two pieces for the inclusive one-jet
and two-jet cross section separately. For the two-jet cross section, it is
essential that the variable $E_T$ is chosen in the way as stated above, namely
that $E_T$ is the transverse momentum of the two largest $E_T$ jets. This
means that the cut-off $y$ is purely technical. It only serves to distinguish
the phase space regions, where the integrations are done analytically with
arbitrary dimensions from those where they are done numerically in four
dimensions. When one compares to experimental data with special kinematical
cuts, $y$ must be chosen sufficiently small in order to make sure that the
cuts on kinematical variables of the final state do not interfere with the
cancellation of the $y$ dependence. \\

For the recombination of two partons $i$ and $j$ with the cone constraint
$R_{i,j}<R$, where 
\begin{equation}
  R_i=\sqrt{(\eta_i-\eta_J)^2+(\phi_i-\phi_J)^2}
\end{equation}
and
$\eta_J$, $\phi_J$ are the rapidity and azimuthal angle of the recombined jet,
we choose the radius $R=1$. In some cases, an ambiguity may occur when two
partons $i$ and $j$ qualify both as two individual jets $i$ and $j$ and as
a combined jet. In this case we only count the combined jet to avoid possible
double counting. It is clear that in NLO the final state may consist of two
or three jets. The three-jet sample consists of all three-body contributions,
which do not fulfill the cone constraint given above. \\

\begin{figure}[p]
 \begin{center}
  {\unitlength1cm
  \begin{picture}(12,8)
   \epsfig{file=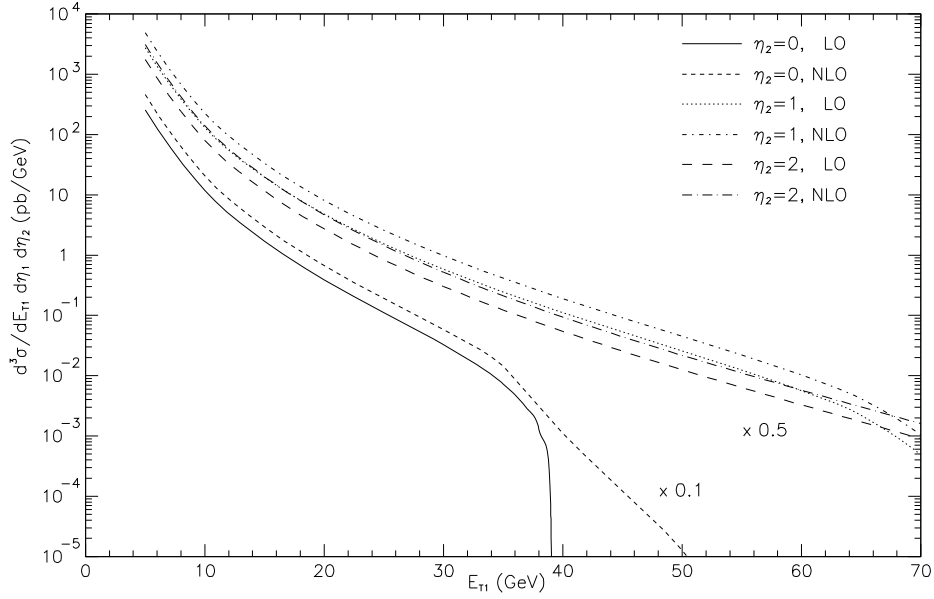,bbllx=520pt,bblly=95pt,bburx=105pt,bbury=710pt,%
           height=12cm,clip=,angle=270}
  \end{picture}}
  \caption[$E_T$-Dependence of Dijet Cross Section for Resolved
           Photoproduction]
          {\label{plot30}{\it Inclusive dijet cross section $\mbox{d}^3
           \sigma/\mbox{d}E_{T_1}\mbox{d}\eta_1\mbox{d}\eta_2$ for resolved
           photons as a function of $E_{T_1}$ for $\eta_1=1$ and three values
           of $\eta_2 =0,1,2$. The cross section for $\eta_2=0$ ($\eta_2=2$)
           is multiplied by 0.1 (0.5).}}
 \end{center}
\end{figure}

First we show results for the resolved inclusive two-jet cross section in
fig.~\ref{plot30}.
\begin{figure}[p]
 \begin{center}
  {\unitlength1cm
  \begin{picture}(12,8)
   \epsfig{file=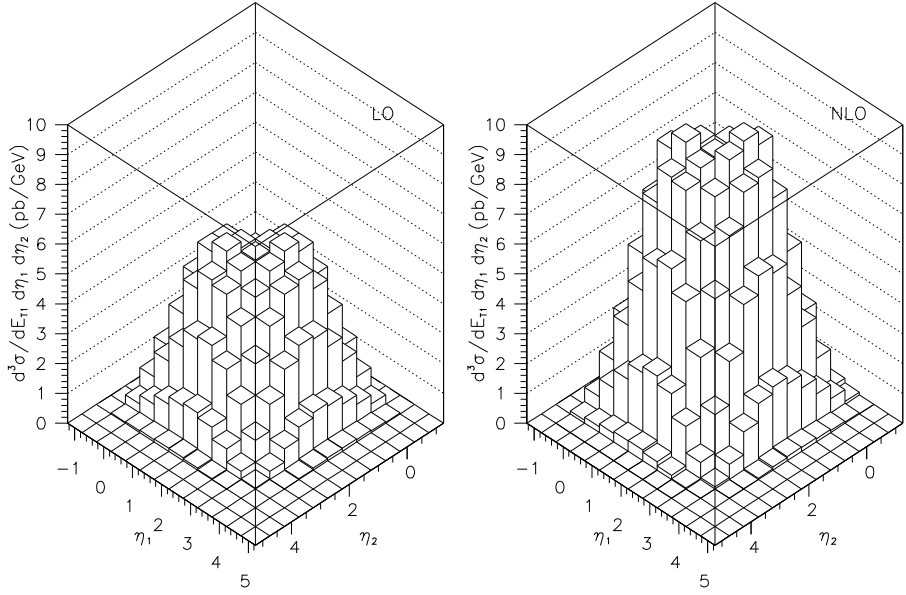,bbllx=520pt,bblly=95pt,bburx=105pt,bbury=710pt,%
           height=12cm,clip=,angle=270}
  \end{picture}}
  \caption[Lego-Plot of Dijet Cross Section for Resolved Photoproduction]
          {\label{plot32}{\it Inclusive dijet cross section $\mbox{d}^3
           \sigma/\mbox{d}E_{T_1}\mbox{d}\eta_1\mbox{d}\eta_2$ at $E_{T_1}=
           20$~GeV for resolved photons, as a function of $\eta_1$ and
           $\eta_2$.
           The LO plot (left) is exactly symmetric, the NLO plot (right) only
           approximately.}}
 \end{center}
\end{figure}
We fixed $\eta_1$ at $\eta_1=1$ and $\eta_2$ at three different
values of $\eta_2=0,1,2$ and plot the cross section as a function of
$E_T=E_{T_1}$. The curves for $\eta_2=0$ and $\eta_2=2$ are rescaled by
factors of 0.1 and 0.5. Since the two-jet cross section is much more exclusive
than the one-jet cross section, it is reduced in magnitude as we expect it.
In fig.~1, we also show the LO cross section as a function of $E_{T_1}$.
The ratio of NLO to LO is 1.7 over the whole $E_{T_1}$ range. In this
comparison, the LO cross section is calculated with the LO subprocesses,
but with the same definition of $\alpha_s$ and NLO structure functions.
For $\eta_2=0$ the NLO and LO differ very much for large $E_{T_1}$. In this
case the third jet in NLO opens up additional phase space, so that $E_{T_1}$
can go up to $E_{T_1} < 50$ GeV. A similar behavior is observed for the
corresponding direct cross section shown in \cite{xxx8}. \\

Next we present the dependence of the resolved cross section on the two
rapidities in form of the three-dimensional lego-plots in fig.~\ref{plot32}.
The equivalent plot for the direct cross section was already given in
\cite{xxx8}. The leading order is shown on the left side and is completely
symmetric in $\eta_1$ and $\eta_2$. The NLO cross section, which is shown
on the right-hand side of fig.~\ref{plot32}, is not symmetric any more due
to the presence of a trigger jet with transverse energy $E_{T_1}$, which
is fixed at $E_{T_1} = 20$ GeV. This can best be seen at the bottom of the
contour plots, where at least one of the two observed jets is far off the
central region. The NLO predictions are considerably larger than the LO
predictions. For the direct cross section, the NLO cross section is only
moderately larger than the LO cross section \cite{xxx8}, where the $k$-factor
is 1.25. \\

\begin{figure}[p]
 \begin{center}
  {\unitlength1cm
  \begin{picture}(12,8)
   \epsfig{file=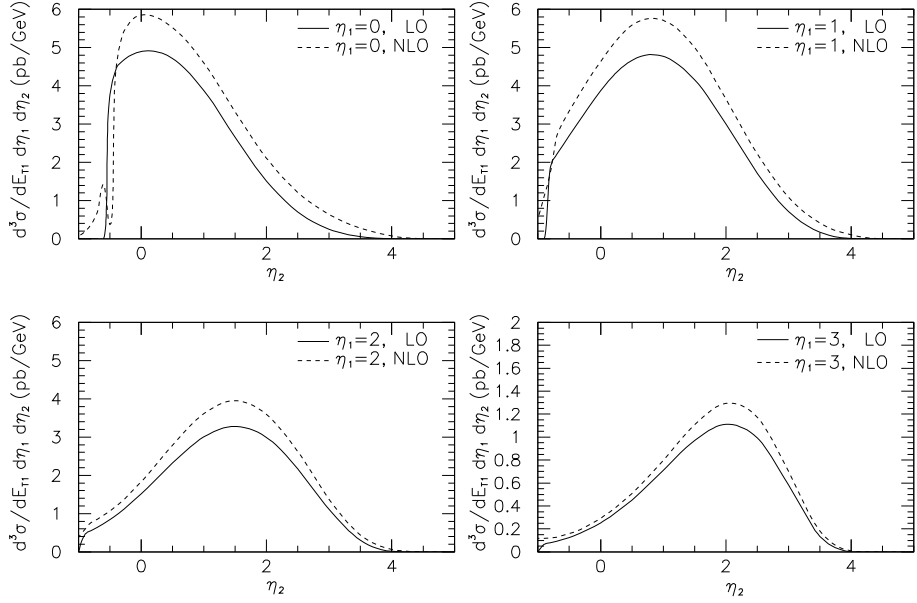,bbllx=520pt,bblly=95pt,bburx=105pt,bbury=710pt,%
           height=12cm,clip=,angle=270}
  \end{picture}}
  \caption[Rapidity Dependence of Dijet Cross Section for Direct
           Photoproduction]
          {\label{plot33}{\it Projections of the LO (full curves) and NLO 
           (dashed curves) triple differential dijet cross section for direct
           photons at $E_{T_1}=20$~GeV and fixed values of $\eta_1=0,~1,~2,$
           and $3$, as a function of $\eta_2$.}}
 \end{center}
\end{figure}

\begin{figure}[p]
 \begin{center}
  {\unitlength1cm
  \begin{picture}(12,8)
   \epsfig{file=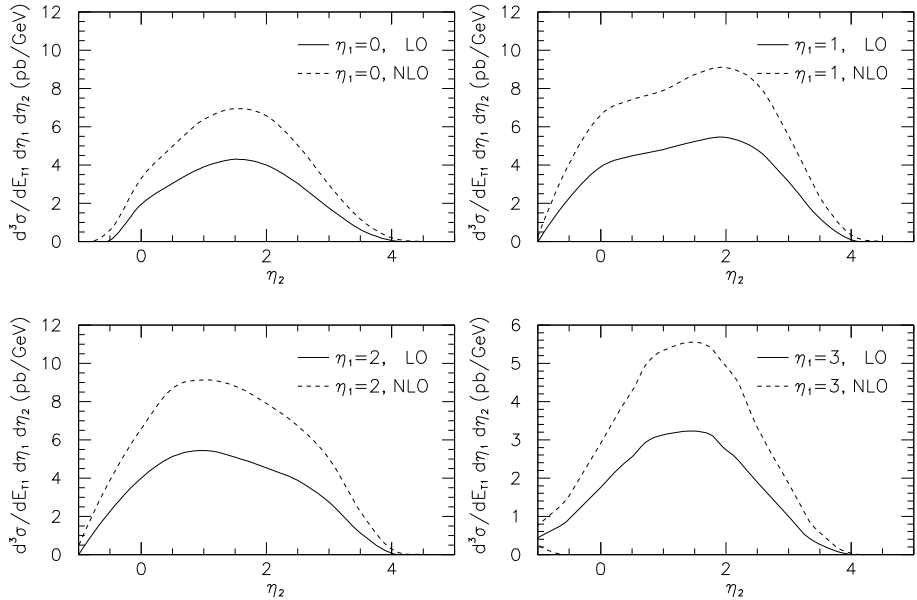,bbllx=520pt,bblly=95pt,bburx=105pt,bbury=710pt,%
           height=12cm,clip=,angle=270}
  \end{picture}}
  \caption[Rapidity Dependence of Dijet Cross Section for Resolved
           Photoproduction]
          {\label{plot34}{\it Projections of the LO (full curves) and NLO 
           (dashed curves) triple differential dijet cross section for resolved
           photons at $E_{T_1}=20$~GeV and fixed values of $\eta_1=0,~1,~2,$
           and $3$, as a function of $\eta_2$.}}
 \end{center}
\end{figure}

This becomes even clearer when we plot the projections of the lego-plots for
fixed $\eta_1=0,~1,~2$ and 3. In fig.~\ref{plot33},
the LO and NLO distributions in $\eta_2$ are shown for the direct
photoproduction and in fig.~\ref{plot34}
for the resolved part. In the direct case it is clearly seen that the second
jet tends to be back-to-back with the first jet, since the maximum always
occurs at $\eta_2\simeq\eta_1$. However, at $\eta_1=3$ this is no longer
possible due to phase space restrictions. The $\eta_2$-distributions for
resolved photons in fig.~\ref{plot34} are considerably broader than in
fig.~\ref{plot33} due to the smearing of the hard cross sections with the
distribution functions of the partons in the photon. The maxima of the
curves are also not so much dominated by kinematics but more by the quark
and gluon structure of the photon in different $x_\gamma$ regimes. They do not
lie at $\eta_2=\eta_1$. Therefore, two-jet rapidity distributions are very
well suited to constrain the photon structure. The $k$-factors range from
1.65 to more than 3 in the very forward region of the proton. The shapes of the
distributions are very similar in LO and NLO. The absolute values make,
however, an important difference. \\

\begin{figure}[p]
 \begin{center}
  {\unitlength1cm
  \begin{picture}(12,8)
   \epsfig{file=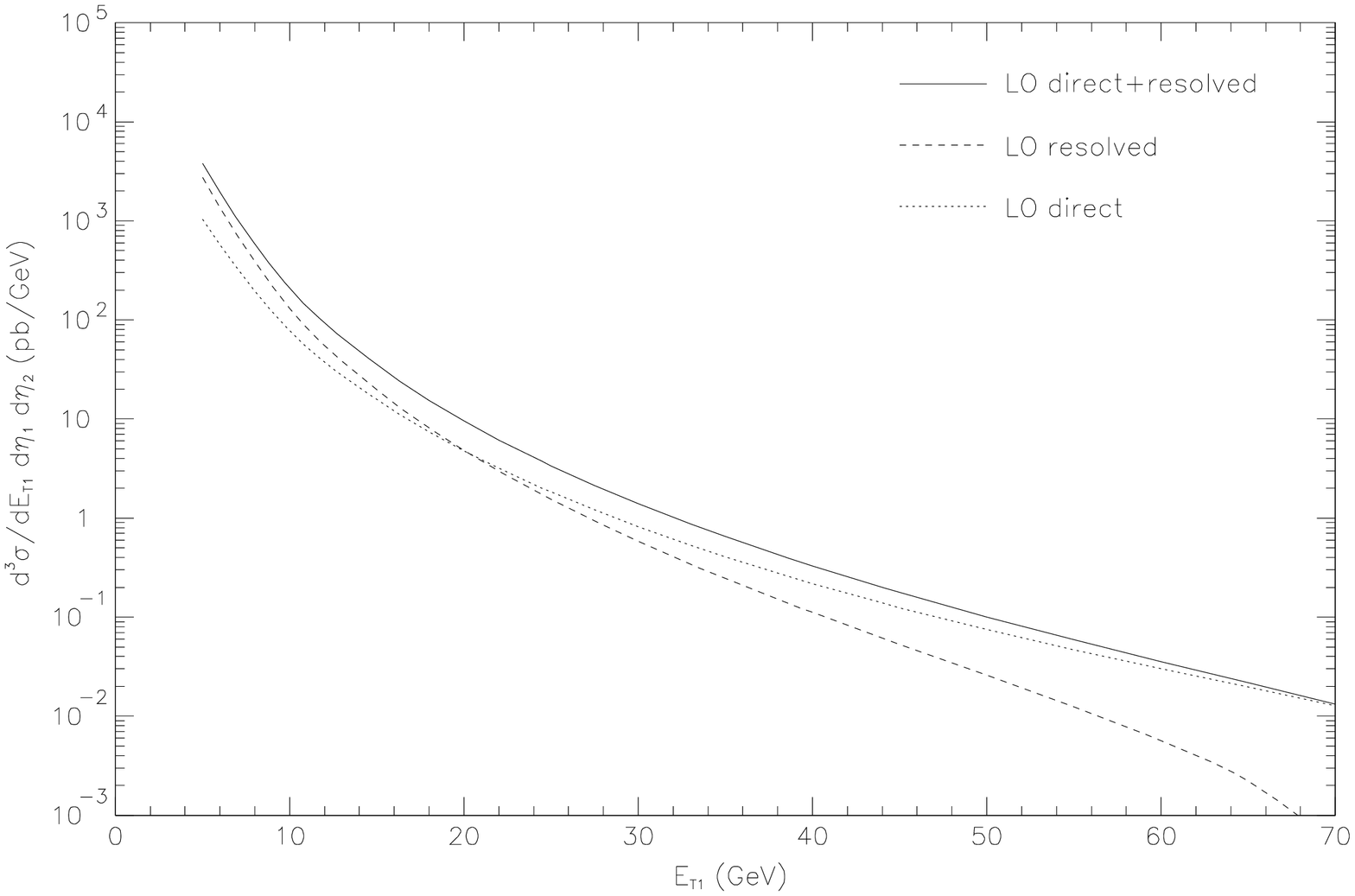,bbllx=520pt,bblly=95pt,bburx=105pt,bbury=710pt,%
           height=12cm,clip=,angle=270}
  \end{picture}}
  \caption[$E_T$-Dependence of Dijet Cross Section for Complete
           Photoproduction in LO]
          {\label{plot37}{\it Inclusive dijet cross section $\mbox{d}^3\sigma
           /\mbox{d}E_{T_1}\mbox{d}\eta_1\mbox{d}\eta_2$ for full
           photoproduction at $\eta_1=\eta_2=1$ as a function of $E_{T_1}$.
           The full curve is the sum of the LO direct (dotted) and LO resolved
           (dashed) contributions.}}
 \end{center}
\end{figure}

\begin{figure}[p]
 \begin{center}
  {\unitlength1cm
  \begin{picture}(12,8)
   \epsfig{file=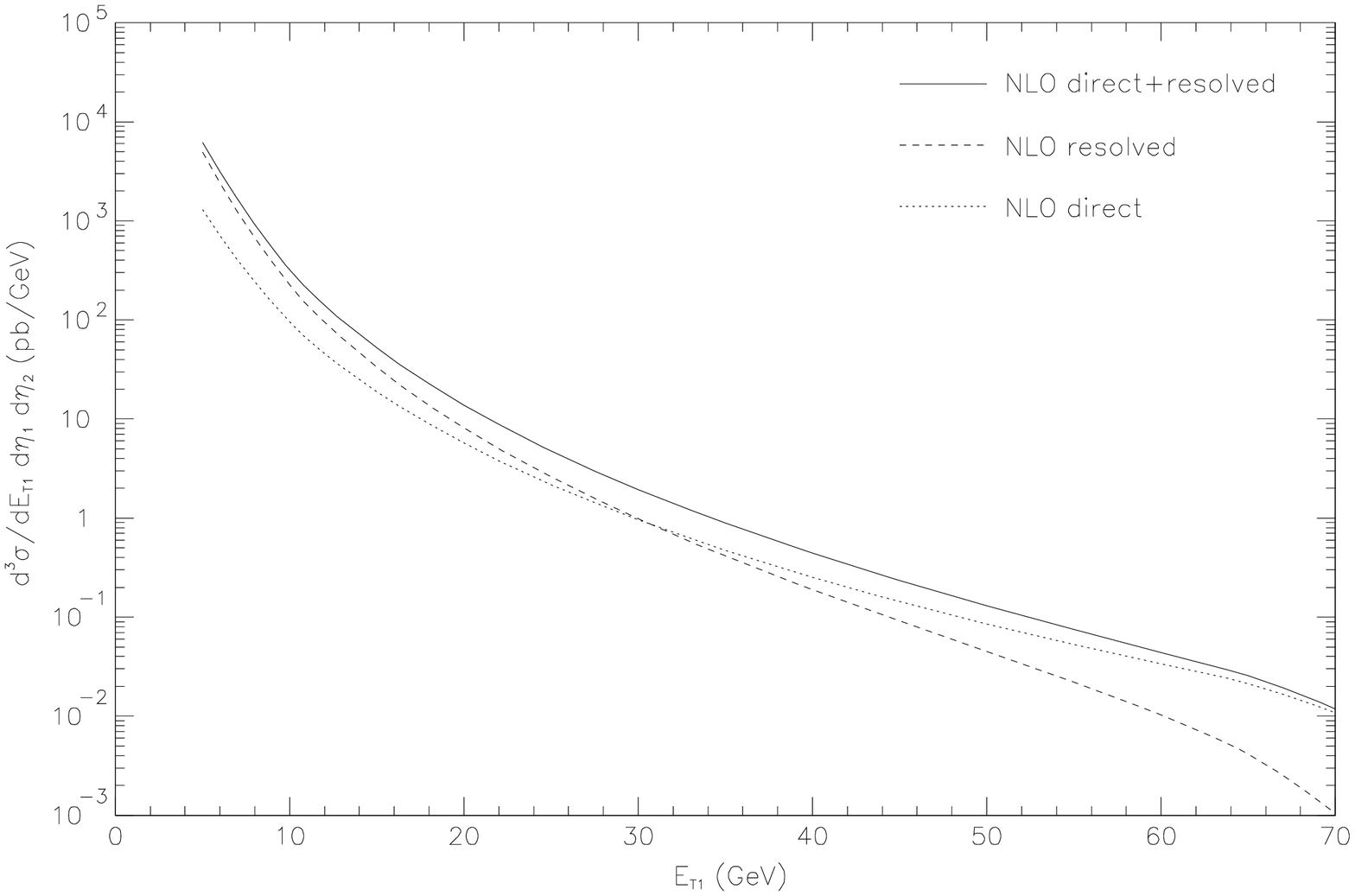,bbllx=520pt,bblly=95pt,bburx=105pt,bbury=710pt,%
           height=12cm,clip=,angle=270}
  \end{picture}}
  \caption[$E_T$-Dependence of Dijet Cross Section for Complete
           Photoproduction in NLO]
          {\label{plot38}{\it Inclusive dijet cross section $\mbox{d}^3\sigma
           /\mbox{d}E_{T_1}\mbox{d}\eta_1\mbox{d}\eta_2$ for full
           photoproduction at $\eta_1=\eta_2=1$ as a function of $E_{T_1}$.
           The full curve is the sum of the NLO direct (dotted) and NLO
           resolved (dashed) contributions.}}
 \end{center}
\end{figure}

Next we show the sum of the direct and resolved cross sections which gives
the physical complete photoproduction result. First we have plotted the
two-jet cross section as a function of the transverse energy $E_{T_1}$ at
$\eta_1=\eta_2=1$. Fig.~\ref{plot37}
gives the LO result, fig.~\ref{plot38}
the NLO result. As we can see, the point where direct and resolved
contributions are equally important is near $E_{T_1}=20$ GeV in leading
order and $E_{T_1}=30$ GeV in next-to-leading order. These crossing points
are somewhat smaller than for the inclusive one-jet cross section at $\eta=1$,
so that direct photons make a stronger impact in dijet production. \\

\begin{figure}[p]
 \begin{center}
  {\unitlength1cm
  \begin{picture}(12,8)
   \epsfig{file=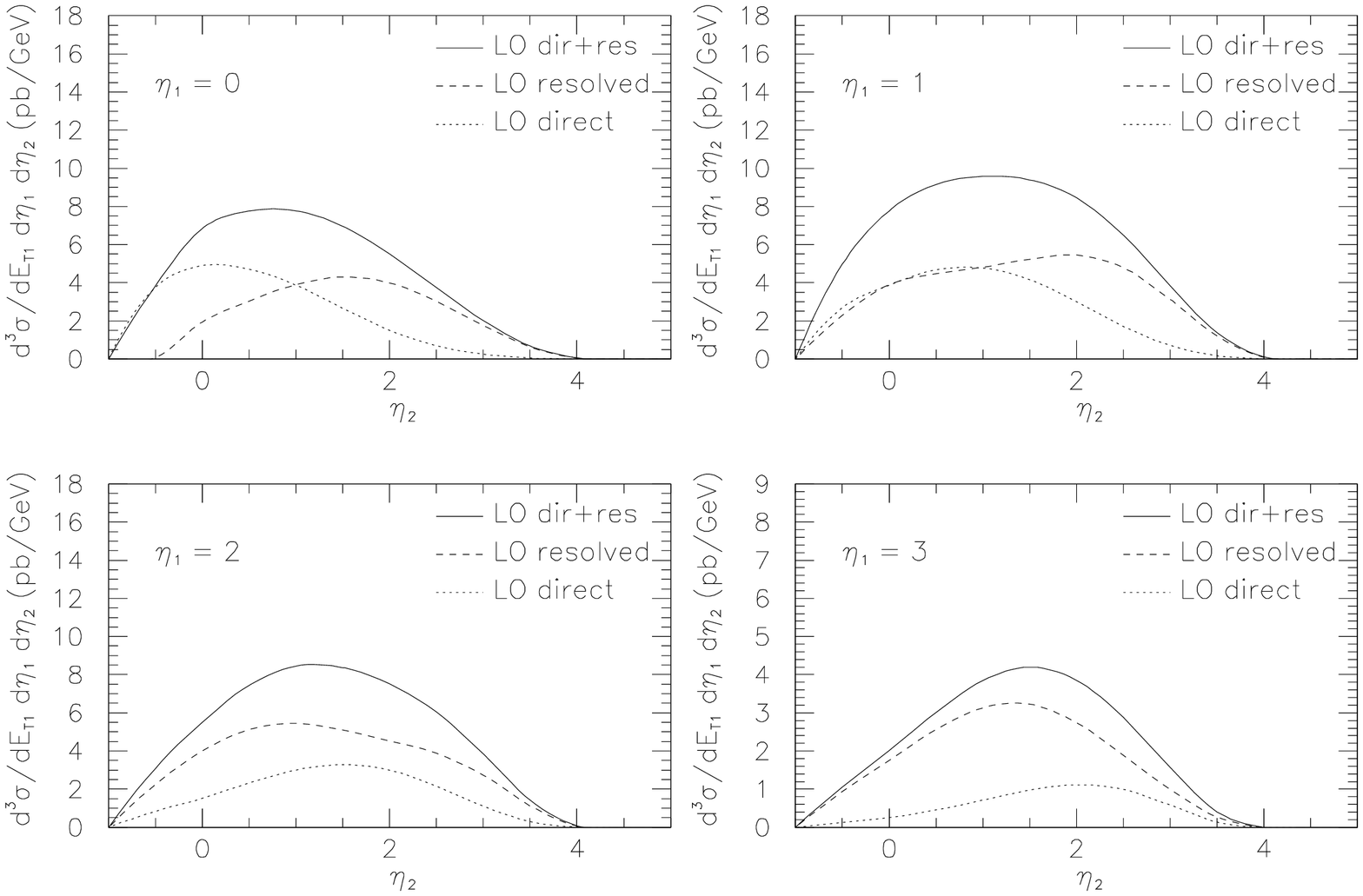,bbllx=520pt,bblly=95pt,bburx=105pt,bbury=710pt,%
           height=12cm,clip=,angle=270}
  \end{picture}}
  \caption[Rapidity Dependence of Dijet Cross Section for Complete
           Photoproduction in LO]
          {\label{plot39}{\it Projections of the complete triple differential
           dijet cross section at $E_{T_1}=20$~GeV and fixed values of
           $\eta_1=0,~1,~2,$ and $3$, as a function of $\eta_2$.
           The full curve is the sum of the LO direct (dotted) and LO
           resolved (dashed) contributions.}}
 \end{center}
\end{figure}

\begin{figure}[p]
 \begin{center}
  {\unitlength1cm
  \begin{picture}(12,8)
   \epsfig{file=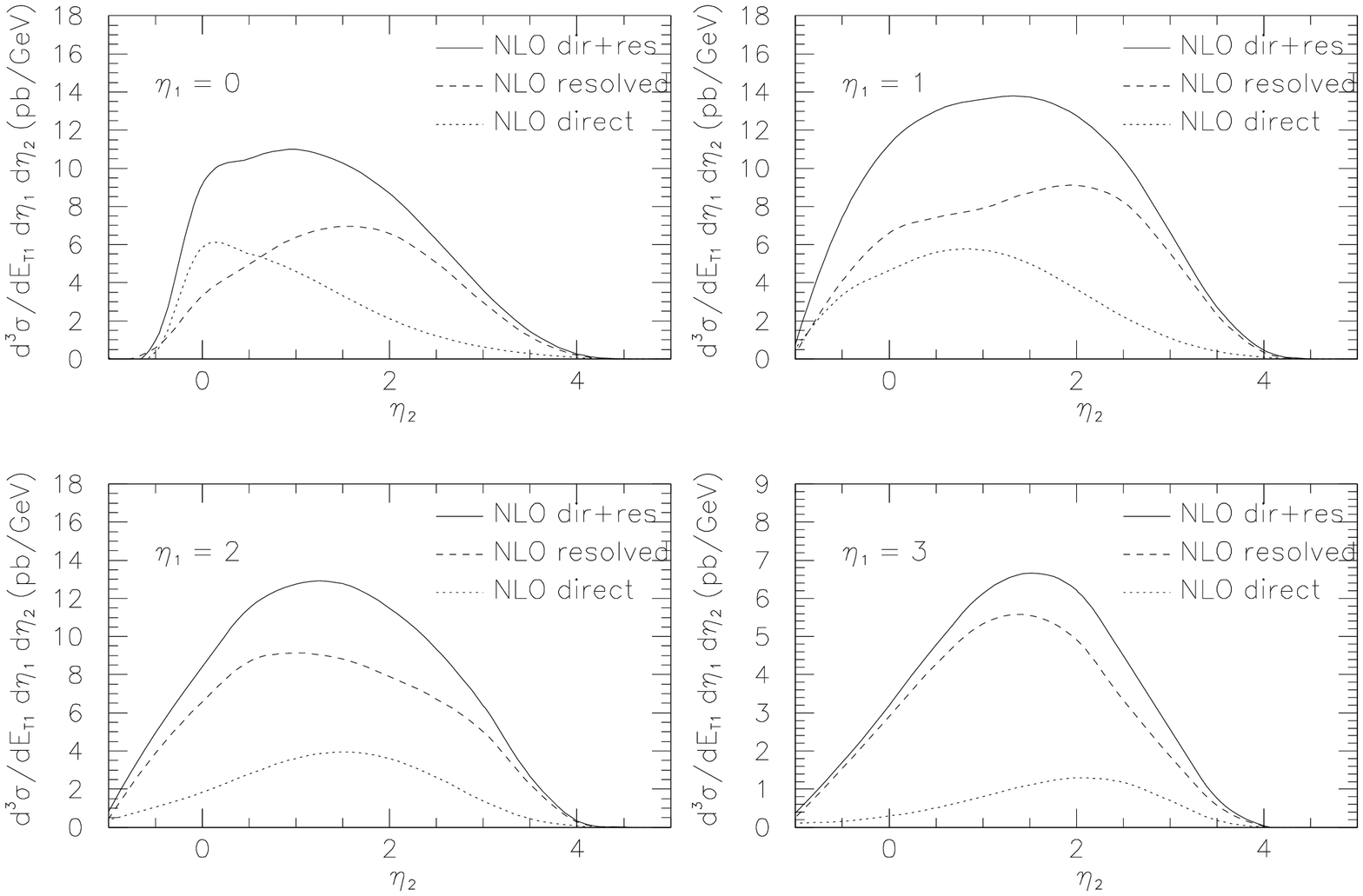,bbllx=520pt,bblly=95pt,bburx=105pt,bbury=710pt,%
           height=12cm,clip=,angle=270}
  \end{picture}}
  \caption[Rapidity Dependence of Dijet Cross Section for Complete
           Photoproduction in NLO]
          {\label{plot40}{\it Projections of the complete triple differential
           dijet cross section at $E_{T_1}=20$~GeV and fixed values of
           $\eta_1=0,~1,~2,$ and $3$, as a function of $\eta_2$.
           The full curve is the sum of the NLO direct (dotted) and NLO
           resolved (dashed) contributions.}}
 \end{center}
\end{figure}

If one plots the complete two-jet cross sections as a function of $\eta_2$,
the different behaviors of direct and resolved photons add up to the full
curves in fig.~\ref{plot39}
(LO) and fig.~\ref{plot40} (NLO).
These plots are best suited to decide in which rapidity regions one can
see best the resolved photon structure. We have already seen that this will
be in situations where the two jets are not back-to-back, e.g.~for $\eta_1=0$
and positive $\eta_2$ in the upper left plots of fig.~\ref{plot39} and
\ref{plot40}. On the other hand, the proton structure can best be studied
with direct photons, when the cross section is not folded with another
distribution. A possibility is at $\eta_1=0$ and negative values of
$\eta_2$. This is especially interesting for the small-$x$ components of the
proton like the gluon and the quark sea. Another interesting observation
is that the relative importance of direct and resolved processes changes
when calculating dijet photoproduction in next-to-leading order ${\cal O}
(\alpha\alpha_s^2)$: resolved processes are much more important at
$E_{T_1}=20$ GeV than one would have guessed from a leading order estimate. \\

For the comparison of our two-jet predictions with experimental data it is
essential that for the theoretical calculations the same jet definitions are
introduced as used in the experimental analysis. First experimental data for
inclusive two-jet production have been published by the ZEUS collaboration
in \cite{xxx7} and \cite{xxx18}. The analysis of these data was based on
the usual cone algorithm with $R = 1$ \cite{xxx12} similar to the cone
algorithm which we used in our NLO calculations presented above. The recent
ZEUS analysis \cite{xxx11} based on the 1994 data taking extends the earlier
analysis \cite{xxx7} based on the 1993 data in several ways. The larger
integrated luminosity utilized in 1994 lead to a reduction of the
statistical errors as well as allowing for the measurement of the cross
section at higher $E_T$, a region, where uncertainties due to non-perturbative
hadronization of partons into jets are expected to be reduced making
the comparison with the NLO predictions more meaningful. Furthermore, the ZEUS
collaboration applied three different jet definitions: two variations of the
cone algorithm \cite{xxx12} are used which they called ``EUCELL'' and
``PUCELL''. These algorithms treat seed finding and jet merging in different
ways. In addition, the $k_T$ cluster algorithm ``KTCLUS'' \cite{xxx19} for
hadron-hadron collisions is used. In \cite{xxx19}, several versions of the
$k_T$ algorithm have been introduced which left some flexibility about how
the stopping condition and the recombination scheme can be implemented.
A version which is particularly suitable to define jets for inclusive jet cross
section measurements was emphasized in \cite{xxx20}. It is also closest to
the cone algorithms and states that two protojets $i$ and $j$ (i.e.~two
partons in the three parton final states) are merged if
\begin{equation}
 \sqrt{(\eta_i-\eta_j)^2+(\phi_i-\phi_j)^2} < 1
\end{equation}
and the recombination of the two protojets is done with the $E_T$ scheme as
is usually also done in the cone algorithms \cite{xxx12}. Compared to the
cone algorithm \cite{xxx12} with $R = 1$, which was applied so far to
generate the results in figs.~1-8, the two partons are merged if they are
less than $R = 1$ apart in $\eta-\phi$ space, whereas in the cone algorithm
the merging condition is on the opening angle between either of the two
partons and the jet center obtained from the $E_T$ recombination equations,
i.e. $R_i < R$ and $R_j < R$ with $R=1$ and $R_i$ defined in equation (5).
Thus this version of the $k_T$ algorithm is very similar to the cone algorithm
usually applied in hadron-hadron and photon-hadron collision. As it is
a cluster algorithm, the ambiguities associated with seed finding and jet
merging are avoided. The same $k_T$ algorithm, denoted by KTCLUS, was used
by the ZEUS collaboration in the analysis of their recent data \cite{xxx11}.\\
\begin{figure}[p]
 \begin{center}
  {\unitlength1cm
  \begin{picture}(12,8)
   \epsfig{file=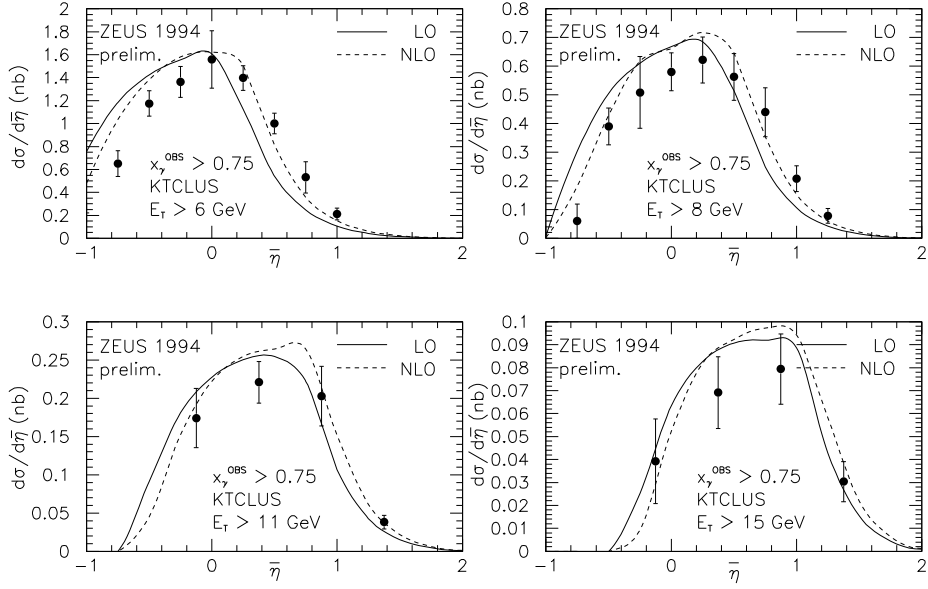,bbllx=520pt,bblly=95pt,bburx=105pt,bbury=710pt,%
           height=12cm,clip=,angle=270}
  \end{picture}}
  \caption{\label{plot9}{\it Inclusive dijet cross section
           d$\sigma/$d$\overline{\eta}$ at $x_{\gamma}^{\rm OBS} > 0.75$
           as a function of $\overline{\eta}$ and integrated over $\eta^{\ast}
           \in[-0.5,0.5]$ and $E_T > 6,~8,~11,$ and $15$ GeV. Our leading and
           next-to-leading order predictions are
           compared to preliminary 1994 data from ZEUS using the
           KTCLUS algorithm.}}
 \end{center}
\end{figure}
\begin{figure}[p]
 \begin{center}
  {\unitlength1cm
  \begin{picture}(12,8)
   \epsfig{file=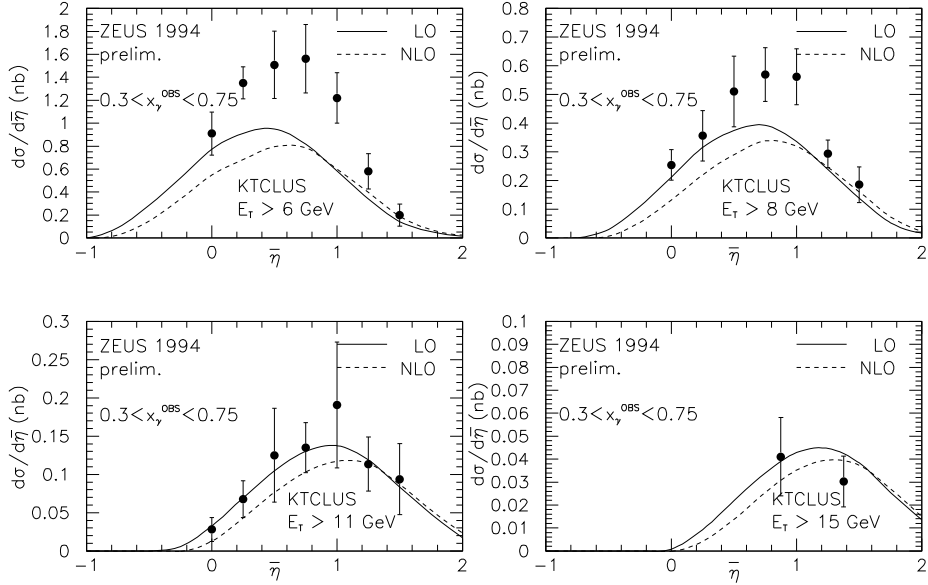,bbllx=520pt,bblly=95pt,bburx=105pt,bbury=710pt,%
           height=12cm,clip=,angle=270}
  \end{picture}}
  \caption{\label{plot10}{\it Inclusive dijet cross section
           d$\sigma/$d$\overline{\eta}$ at $x_{\gamma}^{\rm OBS}\in[0.3,0.75]$
           as a function of $\overline{\eta}$ and integrated over $\eta^{\ast}
           \in[-0.5,0.5]$ and $E_T > 6,~8,~11,$ and $15$ GeV. Our leading and
           next-to-leading order predictions are
           compared to preliminary 1994 data from ZEUS using the
           KTCLUS algorithm.}}
 \end{center}
\end{figure}

Except for the different jet definition, the cross section to be compared
to the ZEUS data is the same as in our earlier work \cite{xxx6}. It is the
inclusive two-jet cross section d$^3\sigma/$d$E_T$d$\overline{\eta}$d$\eta
^{\ast}$, where $\overline{\eta}=\frac{1}{2}(\eta_1+\eta_2)$ is the
average rapidity of the two observed jets and $E_T$ is the transverse
energy of the ``trigger'' jet. This cross section is integrated over the
difference in rapidity, $|\eta^{\ast}| < 0.5$, and integrated over
$E_{T_1},E_{T_2}> E_T^{\min}$ with varying $E_T^{\min}$. The additional
restrictions on the individual rapidities $\eta_1$ and $\eta_2$ used in the
ZEUS analysis have not been implemented since they only modify the cross
sections at $\overline{\eta}\simeq 2$. As the experimental
constraint on the transverse energies of both jets is not infrared safe
in NLO, we allow the second jet to have a transverse energy less than
$E_T^{\min}$ if the third unobserved jet is soft, i.e.~has a transverse
energy of less than 1 GeV ($E_{T_3} < 1$ GeV). Through this procedure we
avoid the dependence of the theoretical prediction on the $y$-cut. In addition,
we separate ``direct'' and ``resolved'' contributions with the
variable
\begin{equation}
 x_{\gamma}^{\rm OBS} = \frac{\sum_i E_{T_i}e^{-\eta_i}}{2x_aE_e},
\end{equation}
which measures the fraction of the photon energy that goes into the production
of the two hardest jets. As in the ZEUS analysis the enriched direct $\gamma$
sample is defined with the cut on $x_{\gamma}^{\rm OBS} > 0.75$. The enriched
resolved $\gamma$ sample is obtained for $0.3 < x_{\gamma}^{\rm OBS} < 0.75$.
The very low $x_{\gamma}^{\rm OBS}$ are excluded in the experimental
analysis since this region is not accounted for by the Monte Carlo routines
necessary to correct for detector effects. Fig.~\ref{plot9}
shows the ``direct'' cross sections d$\sigma/$d$\overline{\eta}$ as a function
of $\overline{\eta}$ in LO and in NLO for $E_T^{\min} = 6,~8,~11$, and 15 GeV
together with the preliminary ZEUS data \cite{xxx11}. The LO and NLO
predictions are very similar. We emphasize that the LO curve is calculated
with the same NLO structure functions as in the NLO calculations, only the
hard scattering parton-parton cross sections are evaluated in LO. The NLO
curves are shifted to larger $\overline{\eta}$ and agree in average better
with the data than the LO curves. The general agreement is quite good
indicating that the GRV choice for the photon parton distributions needed for
the resolved contribution is quite reasonable in the high $x_{\gamma}$ region.
In our previous publication \cite{xxx6}, we compared the corresponding 1993
ZEUS data to our prediction, which then only included the LO resolved
contributions for the GRV photon parton densities. \\

The ``resolved'' photon two-jet cross section, where $0.3 < x_{\gamma}^{\rm
OBS} < 0.75$, is compared to the ZEUS data in fig.~\ref{plot10}
for the same $E_T^{\min}$ values. Here the agreement is satisfactory only
with the data in the high $E_T$ regions $E_T^{\min} = 11$ and 15 GeV. At
the lower $E_T$, the experimental cross section is larger than the NLO
prediction. The disagreement increases with decreasing $E_T^{\min}$. We
attribute this difference between theory and experimental data to additional
contributions due to multiple interactions with the remnant jet not accounted
for by our NLO predictions. As one would expect that these effects diminish
with increasing $E_T$, the agreement between theory and experiment
improves in the large $E_T$ region. We also studied the uncertainty coming
from the insufficiently constrained gluon in the photon by recalculating the
resolved cross section with twice the GRV gluon distribution. At low $E_T$,
the cross section is enhanced by 33\% thus reducing the discrepancy, whereas
at large $E_T$ it is only enhanced by 20\%. With more accurate data at $E_T >
17$ GeV, such a gluon distribution could be excluded. In addition to the
experimental errors
shown, which include systematic and statistical errors added in quadrature,
there is a systematic uncertainty arising from the uncertainty in the
calorimeter energy scale, which is highly correlated between bins and is
therefore excluded from the systematic errors shown in figs.~\ref{plot9} and
\ref{plot10}. This uncertainty is largest for the ``resolved'' cross section
at $E_T^{\min} = 6$ GeV and leads to a $\pm 0.5$ nb uncertainty in the
cross section near $\overline{\eta}\simeq 0.6$ \cite{xxx11}. In the ``direct''
cross section, this energy scale uncertainty is smaller by a factor of two
\cite{xxx11}. \\

In fig.~\ref{plot10}, we observe that the NLO two-jet cross sections are
somewhat smaller than the LO cross sections, which are included only to
see the effect of the NLO corrections. These LO cross sections are independent
of the jet definition, i.e.~are always the same for the cluster algorithm or
a cone algorithm with any cone radius. They should not be compared to the
experimental data since it is known that the two-jet cross sections depend
on the jet definitions. The choice of the jet definition has an effect of
about 25-30\% in both data and theory in the ``direct'' cross section and
of about 50\% for the ``resolved'' cross section \cite{xxx11,xxx21}.

\section{Summary}

Differential cross sections d$^3\sigma$/d$E_T$d$\eta_1$d$\eta_2$ have been
calculated in NLO for direct and resolved photoproduction. Infrared and
collinear singularities are cancelled with the phase space slicing method
using an invariant mass cut-off. With this method we are able to incorporate
various cuts on the final state as used in the analysis of experimental
data and to perform calculations for different choices of jet algorithms.
Numerical results for the two-jet inclusive cross sections at HERA have
been presented employing the usual Snowmass cone algorithm for the jet
definition. For a cone radius of $R=1$, the NLO corrections lead to an
increase of the order of 70\% compared to the LO prediction in the resolved
contribution and of the order of 20\% in the direct case already presented
earlier \cite{xxx8}.\\

Using a particular version of the $k_T$ cluster algorithm \cite{xxx20},
we calculated the cross section d$\sigma$/d$\overline{\eta}$ with cuts on
$x_{\gamma}^{\rm OBS}$ to separate ``direct'' and ``resolved'' contributions
as in the analysis of the ZEUS data \cite{xxx7,xxx11,xxx18}. We find
fairly good agreement with recent ZEUS data \cite{xxx11} for the enriched
direct $\gamma$ sample. For the enriched resolved $\gamma$ sample, the
agreement is good in the larger $E_T$ region, $E_T^{\min} \geq 11$ GeV. For
smaller $E_T$ ($E_T^{\min} = 6$ and 8 GeV), we find a discrepancy which
we attribute to additional multi-parton interactions \cite{xxx11} not
incorporated in the theoretical calculations.

\section*{Acknowledgements}
It is a pleasure to thank J. Butterworth and L. Feld for valuable discussions
on the ZEUS dijet analysis.

\end{document}